# The Tongue as an Excitable Medium

Gabriel Seiden* and Sofia Curland

Department of Earth and Planetary Sciences, Weizmann Institute of Science, Rehovot 76100, Israel

**Geographic tongue[1-7] (GT) is a benign condition affecting approximately 2% of the population[1], whereby the papillae covering the upper part of the tongue are lost due to a slowly expanding inflammation. The resultant dynamical appearance of the tongue has striking similarities with well known phenomena observed in excitable media, such as forest fires[8,9], cardiac dynamics[10-14] and chemically-driven reaction-diffusion systems[15-18]. Here we explore the dynamics associated with GT from a dynamical systems perspective, utilizing cellular automata simulations. We emphasize similarities with other excitable systems as well as unique features observed in GT. Our results shed light on the evolution of the inflammation and contribute to the classification of the severity of the condition, based on the characteristic patterns observed in GT patients.**

Some of the most important organs in the human body incorporate microscopic coupled elements possessing a property of excitability. The resultant network usually performs a specific function, crucial for the successful operation of the organ. Under certain conditions, however, this excitable medium can experience unwanted and in some instances harmful dynamics. Probably the most well known example is the heart muscle, which under normal conditions contracts periodically due to the propagation of electrical waves resulting in blood being pumped to the different parts of the body. Irregular heart rhythm (cardiac arrhythmias) caused either by abnormal initiation of impulses[13] or due to the occurrence of spiral waves[11,12] may be life-threatening. Yet another example is the cerebral cortex, which under normal conditions is responsible for important functions such as conciseness and memory, but may experience dangerous electrophysiological waves (i.e., *spreading depression* waves[19,20]).

The human tongue is vital for tasting, mastication and speech. With respect to the former function, a thin layer (epithelium) consisting of four types of tiny papillae covers the upper part of the tongue (see Fig. 1a). Taste buds are located on three of the four papillae. The condition of geographic tongue (also known as *benign migratory glossitis*) is associated with the loss of the fourth papilla, called filiform papilla. Filiform papillae (FP) are mainly distributed in the anterior two-thirds (oral part) of the tongue and do not contain taste buds. While the actual cause of GT is yet unknown, possible related conditions include asthma[5], psychological stress[4] and eczema[5].

---

* Email: gabriel.seiden@weizmann.ac.il

Figure 1b depicts a typical scene observed as the ulcer-like regions depleted of FP expand on the upper part (dorsum) of the tongue. The expansion of these lesions throughout the tongue and subsequent healing typically lasts several days, although considerably longer periods have been reported[6]. The border between the unaffected part and the inflamed regions may appear as an elevated whitish rim (see Fig. 1c). Note the gradual change of color within the lesion, from red (or dark pink) adjacent to the rim to bright pink away from the borderline. Apart from the expanding oblate pattern (Fig.1b) typically observed in patients with GT, other patterns, such as spirals (see Fig.1d) are also observed.

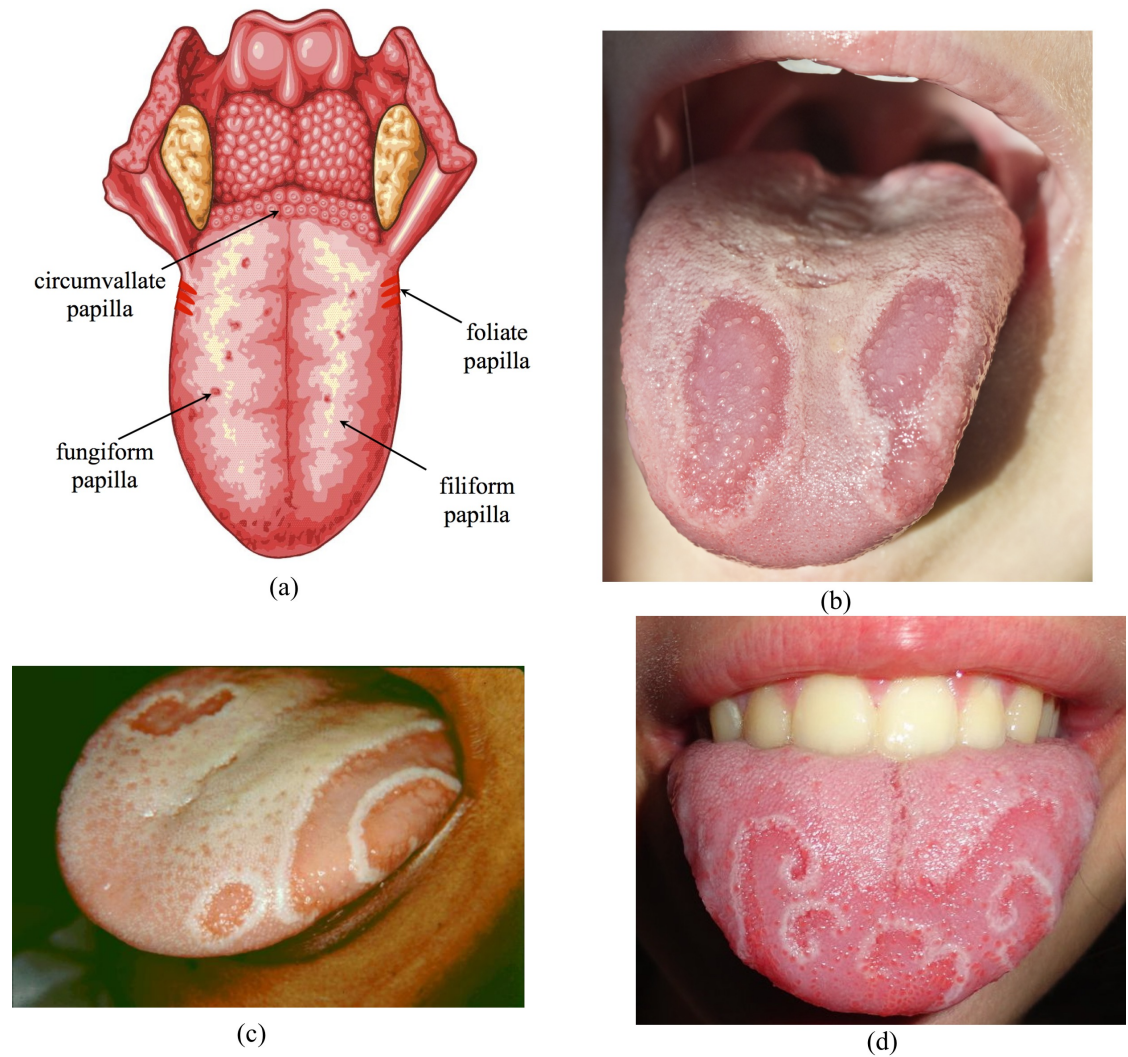

Figure 1: (a) Illustration of dorsum of the human tongue, showing the four types of papillae. Geographic tongue is associated with the loss of filiform papillae. (b) Oblate lesions on the dorsum of the tongue. The tiny 'hairs' covering the unaffected regions are the filiform papillae. Note the gradual change in color within the lesions, from dark pink adjacent to the margin of the lesions to light pink away from the margin. (c) A large lesion on the lateral part of the tongue with a typical whitish rim that marks the border between the affected and unaffected regions. Note the whitish rim *within* the affected area. (d) Different patterns formed by open-ended rims evolving within the expanding lesion. Note in particular the spiral pattern on the left. Figures reproduced with permission[†].

---

[†] (a) Oguz Aral/Shutterstock.com (b) Angel Simon/Shutterstock.com
(c) doctorspiller.com (d) Martanopue/CC BY-SA (Wikipedia).

In a similar manner to other excitable media, GT dynamics evolves between three main states: a rest state (healed epithelium), an excited state (highly inflamed epithelium) and a recovering state (healing epithelium). A local region initially at rest will be excited depending on the overall state of excitability in its vicinity. Once excited, the affected region enters a relatively long recovery state in which its threshold of excitability is higher than its value in the rest state. The resultant characteristic features of excitable media are their sensitivity to small perturbations and their ability to sustain the propagation of solitary as well as periodic waves.

From a theoretical point of view, the dynamics pertaining to excitable media can be modeled through the Oregonator model[17,18,21], first proposed for the Belousov-Zhabotinski (BZ) reaction:

$$\partial_t u = \varepsilon^{-1} f(u, v) + D_u \nabla^2 u \qquad (1a)$$

$$\partial_t v = g(u, v) + D_v \nabla^2 v \qquad (1b)$$

In the above, $u$ and $v$ are the excitation and recovery variables, respectively, $f$ and $g$ represent the corresponding growth rates, and $D_u$ and $D_v$ are the corresponding diffusion constants. The small parameter $\varepsilon$ represents the ratio of the excitation and recovery time scales. Insight into the evolution of the system from an arbitrary state can be gained by considering the nullclines $f = 0$, $g = 0$ in the phase plane spanned by $u$ and $v$ (see Fig. 2a).

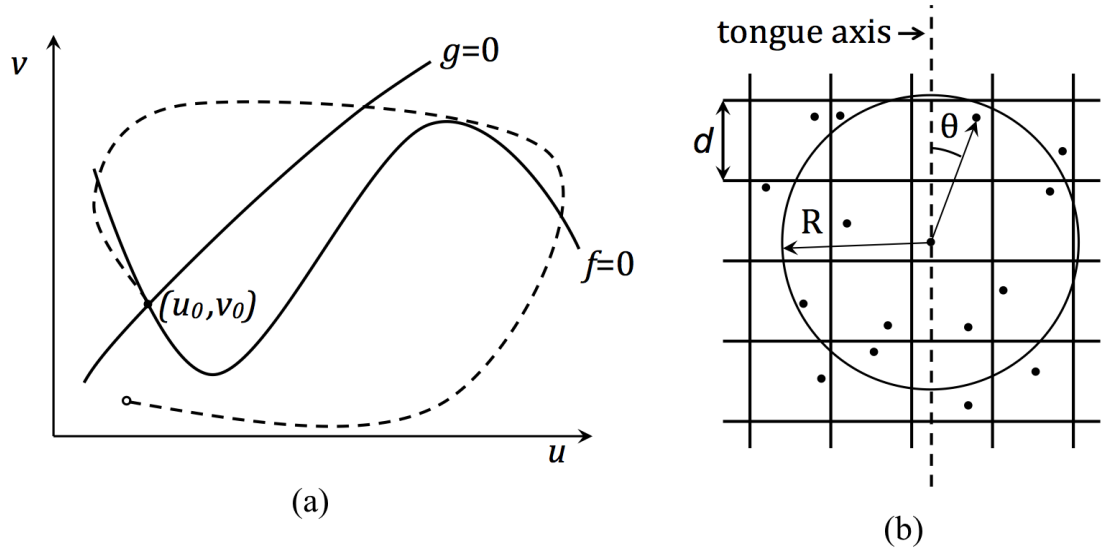

Figure 2: (a) A typical phase plane pertaining to an excitable medium depicting the nullclines corresponding to Eq. 1. The point $(u_0, v_0)$ is a global stable fixed point. A typical initial condition (open circle) results in a lengthy route in phase space until the system eventually reaches its stable point. (b) Schematic describing the grid used in the CA simulations. Grid points are randomly distributed within each square cell of side $d$. A circle of radius $R$ marks the neighborhood of a point in the grid. The angle $\theta$, measured with respect to the tongue axis, allows for the inclusion of anisotropy in the model (see Eq. 2).

While different excitable media share fundamental similarities each system also has unique features, originating either from externally imposed constrains or from intrinsic composition, which affect the dynamics observed. Forest fires, for example, are strongly affected by winds, typically resulting in an anisotropic propagation of the fire-front. The heart muscle, on the other hand, contains specific pacemaker cells, which are responsible for the triggering of periodic electrical pulses. A detailed examination of GT dynamics reveals that both internal composition and external conditions pertaining to the tongue may have important consequences on the dynamics observed.

It has already been noted that the propagation of an inflamed region on the dorsum of the tongue is typically anisotropic[7]. This can be easily observed by examining Fig. 1b where the patterns are oblate (with the long axis approximately parallel to the tongue symmetry axis) rather than circular, as well as in Fig. 1d in which the large circular arc (i.e., the rim between the recovering region and the unaffected region) is stretched along the same axis. This anisotropy might be related to the fact that the thin stratum immediately below the tongue surface is made of longitudinal muscles[22] (the *superior longitudinal* intrinsic muscle) oriented parallel to the tongue axis. A similar situation occurs in cardiac dynamics, where due to anisotropy in the epicardial muscle wave propagation is anisotropic[11].

Apart from its unique composition the tongue is also exposed to external conditions, such as large temperature variations associated with different foods, which may have an affect on the dynamics observed in GT patients. A concrete example of the influence of external conditions experienced by the tongue was observed in a baby boy, diagnosed with GT at the age of one year. During teething we noticed on multiple occasions that the inflamed, FP depleted regions emerged on the tongue's edge, adjacent to a growing tooth, implying that the continuous rubbing of the tongue against the gum may trigger the phenomenon in GT patients.

In order to further examine GT dynamics we utilize cellular automata. Cellular automata (CA) represent a more tangible and far less time-consuming numerical approach than directly integrating the set (1). Here, the excitable medium consists of an array of cells with a set of rules, which determine the state of a given cell at a given time step based on its state and the state of its neighbors in the preceding time step. Different CA schemes have been used to model excitable media dynamics[23-25]. Particularly worth noting is the CA proposed by Markus & Hess[23](1990). Apart from accounting for coupling and diffusion mechanisms, their CA solves the inherent anisotropy stemming from modeling over a periodic lattice through randomly distributing the grid points while maintaining grid uniformity.

In the CA proposed by Markus & Hess the state of a particular cell is represented by one variable $S$, which acquires non-negative integer values. The rest state is denoted $S = 0$, the excited state $n+1$ and the recovery state by the intermediate integer values,

$0 < S < n + 1$. A cell initially in the rest state is excited if the number of excited cells in its neighborhood (defined as a circle of radius $R$; see Fig. 2b) exceeds a given threshold, $m_0$. The threshold for excitability of cells initially in the recovery state is linearly dependent on the state of excitability, $m_0 + pS$ (here $p \geq 0$ is the linear coefficient). However, a recovering cell can only be excited if $S \leq S_{max}$, ensuring a minimal remission period. Diffusion transfer is accounted for by defining an intermediate state, which is integrated in order to acquire the state at the next time step.

The above CA can be modified to account for anisotropy in the expansion of the FP depleted regions. This can be done by generalizing the excitability condition in the following way:

$$\frac{1}{1+a} \sum_i [1 + a \cdot \cos(\theta_i)] \geq m_0 + p \cdot S \qquad (2)$$

where the summation is over all cells within a circle of radius $R$, $\theta_i$ is the angle of the $i^{th}$ cell with respect to the tongue axis (see Fig. 2b) and $a$ represents the weight of the anisotropy. Note that for the isotropic case ($a = 0$) the expression on the right hand side reduces to the number of excited cells. Further details regarding the CA used may be found in the methods section.

Figure 3 depicts CA simulation results of two evolving lesions as they spread from initial circular spots throughout the oral part of the tongue. In the figure, excited regions are marked red and healed (unaffected) regions white. As the lesions expand, they acquire an oblate shape due to the anisotropy term in (2). The lesions merge upon contact ($t=10\tau$; $\tau$ being the numerical time step) and continue spreading until the entire oral part is affected ($t = 30\tau$). The subsequent recovery of the epithelium is observed at $t = 70\tau$.

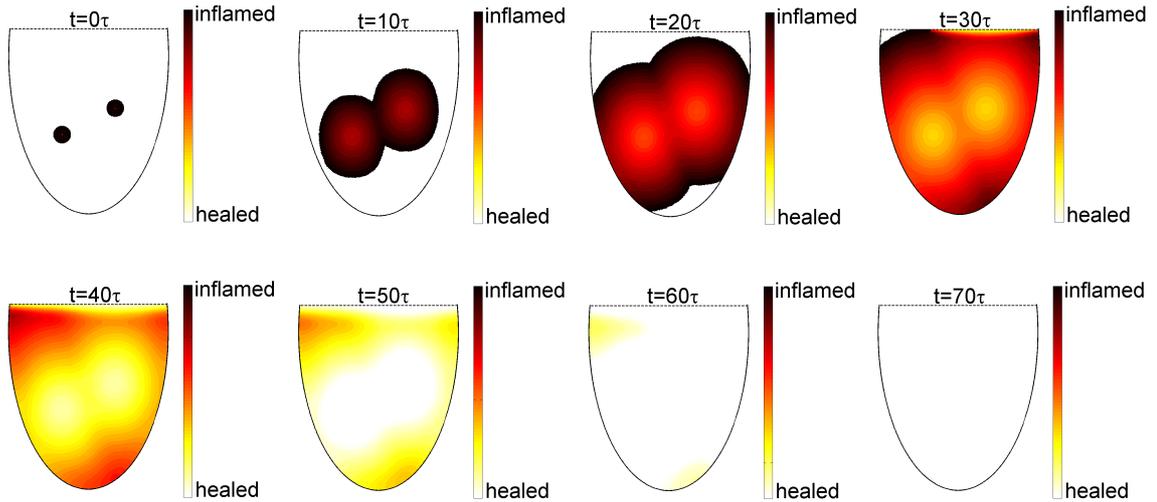

Figure 3: Cellular automaton results depicting the evolution of two lesions from initial circular spots ($t = 0$). The oblate shape of the expanding lesions ($t=10\tau$) is a manifestation of the anisotropy term in Eq. 2. The lesions merge upon contact as they expand throughout tongue. Full recovery of the epithelium is seen at $t = 70\tau$. The oral part of the tongue is modeled as two-thirds of an ellipse (major axis = 600 pix, minor axis = 300 pix). The CA parameters used are: $n = 40$, $S_{max} = 30$, $R/d = 8$, $m_0 = 5$, $p = 1$ and $a = 1$.

An intriguing aspect of GT dynamics is the occurrence of spiral patterns. As depicted in Fig. 1d, spiral patterns in GT tend to evolve in the slowly recovering regions, with no apparent fully recovered FP in their vicinity. Note that as in the case of a closed rim marking the border of an expanding oblate lesion (e.g., Fig. 1b), the wake of the propagating front within the recovering region is marked by an inflamed (reddish) epithelium (Fig. 1c and Fig. 1d). A similar situation was observed in an investigation of spiral wave propagation in an isolated chicken retina[26]. In other cases, however, such as the heart muscle and BZ reaction, the evolution of a spiral wave typically involves the whole spectrum of states (except possibly at the spiral core[11,27]). Spiral wave propagation in recovering regions can occur providing that three conditions are met. The first condition is that a broken excitation front (i.e., open-ended front) is present in the recovering region, the second is that the excitability threshold for recovering states is relatively low and the third condition is that the recovery time is larger than the spiral rotation period. With respect to the first condition, it should be noted that the mechanism of spiral formation from a broken front propagating into an unaffected region is well documented[18,21]. The broken front can occur due to local inhomogeneity caused either by external intervention[26] or by non-uniform medium properties[23]. In the case of GT, broken fronts observed within the recovering region might be the cause of the merging process of two expanding oblate lesions, in which the whitish rim does not fully disintegrate.

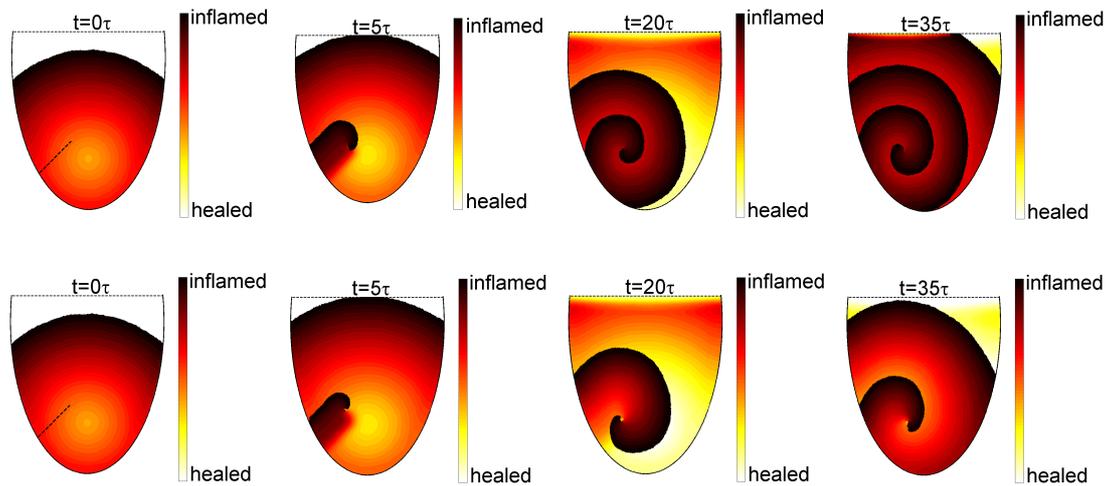

Figure 4: Cellular automaton results depicting the evolution of a broken front initiated within a recovering region, for two different excitation scenarios. The dashed line at $t = 0$ marks the location of the initial front. The top row shows the evolution for a relatively low excitation threshold ($m_0 = 5, p = 1$). The open-ended front curls into recovering regions and forms a spiral, which expands throughout the tongue. The lower row depicts a case with higher excitation values ($m_0 = 5, p = 4$). The large excitation barrier slows the transformation of the front into a spiral and causes the distance between spiral turnings to grow. The CA grid used is as in Fig. 4. The other CA parameters used in both cases are: $n = 40$, $S_{max} = 32$, $R/d = 8$ and $a = 0.1$.

Figure 4 shows CA results pertaining to the evolution of a broken front in a recovering region affected by a preceding oblate front propagation, for two different excitation threshold scenarios. The upper row shows the evolution of the broken front for relatively low excitation threshold ($m_0 = 5, p = 1$). Once formed ($t = 0$), the broken front starts curling at the open end (t=5τ), initiating the formation of a spiral. Note that as the spiral evolves there is no full recovery of the tongue epithelium (t=20τ).

This is due to a relatively low excitability threshold, which allows the spiral to curl and expand into recovering regions. The spiral expands as it grows and eventually covers the whole tongue (t=35τ). Increasing the threshold parameters ($m_0 = 5$, $p = 4$) hinders the open-ended tip from curling into newly affected regions, forcing it to migrate towards regions with lower excitability threshold, where it can evolve into a spiral (lower row of Fig. 4). Note also the relatively large distance between spiral turnings in the latter case ($t=35\tau$).

A comparison between the long time dynamics pertaining to oblate pattern propagation and spiral propagation, as depicted in Fig. 3 and Fig. 4, respectively, sheds light on the question of the severity of the GT condition. While the propagation of oblate patterns results in the whole tongue being gradually affected and subsequently healed, the propagation of spiral patterns involves a continuous, self-sustaining excitation of recovering regions, implying a more acute condition. It is interesting to note that spirals play a negative role also in the heart muscle, where their occurrence is associated with cardiac arrhythmias[11,12]. It is hoped that this insight, gained from a dynamical systems approach to GT dynamics, might assist physicians in the assessment of the severity of GT condition.

**Methods**

The CA used in our investigation is based on a previously published model[23]. In the original CA, the authors used an intermediate state in order to obtain diffusion transport through averaging. The authors did not use averaging for specific cases in which a cell was in one of three states: $S = 0$, $S = 1$ and $S = n+1$, in order to avoid numerical instabilities in the vicinity of the advancing front. As in our simulations there is special emphasis on front propagation in recovering regions, we generalized the above exception in the following manner. Averaging was excluded for cells in the vicinity of the front, for which the standard deviation calculated in their neighborhood exceeded a given value. This ensured that numerical instabilities in the location of the expanding front were avoided regardless of whether the propagation was in fully recovered or partially recovered regions.

**Acknowledgements**
We would like to thank E. Bodenschatz and B. Pokroy for reading and commenting on the manuscript.

**Author contributions**
GS conceived the investigation, wrote the CA codes, ran the simulations and wrote the manuscript. SC carried out direct observations of GT evolution, read and commented on the manuscript.